\documentclass[conference]{IEEEtran}

\usepackage{textcomp}

\usepackage[dvipsnames]{xcolor}

\usepackage{lipsum}

\usepackage{mathtools}
\usepackage{mathrsfs}
\usepackage{amssymb}
\usepackage{nccmath}
\usepackage{gensymb}
\usepackage{nicefrac}
\usepackage{bm} 

\usepackage{siunitx}
\DeclareSIUnit{\dBm}{dBm}

\usepackage{array}
\usepackage{booktabs}
\usepackage{makecell}

\usepackage{tikz}
\usetikzlibrary{
	arrows,             
	arrows.meta,
	automata,           
	backgrounds,        
	calc,               
	chains,
	decorations.pathmorphing,
	decorations.markings,
	fit,                
	intersections,      
	snakes,
	shapes,             
	patterns,
	matrix,             
	mindmap,            
	shadows,
	spy,
	petri,
	plotmarks,
	positioning,
	pgfplots.groupplots,
}

\usepackage{circuitikz}
\usepackage{pgfplots} 
\pgfplotsset{compat=1.14}
\usepgfplotslibrary{fillbetween}
\tikzstyle{every picture}+=[font=\sffamily]

\pgfplotsset{
    legend image with text/.style={
        legend image code/.code={%
            \node[anchor=center] at (0.3cm,0cm) {#1};
        }
    },
}

\definecolor{thesisCreddeep}{RGB}{178,24,43}
\definecolor{thesisCredlight}{RGB}{252,78,42}
\definecolor{thesisCredbright}{RGB}{240,24,43}
\definecolor{thesisCredtcl}{RGB}{252, 172, 172}

\definecolor{thesisCteal}{RGB}{166,189,219}

\definecolor{thesisCorangedeep}{RGB}{236,112,20}
\definecolor{thesisCorangelight1}{RGB}{214,197,115}
\definecolor{thesisCorangelight2}{RGB}{244,196,69}

\definecolor{thesisCpurpledark}{RGB}{122,1,119}

\definecolor{thesisCgreenlight}{RGB}{168,221,181}
\definecolor{thesisCgreendeep}{RGB}{39,143,54}
\definecolor{thesisCgreenbright}{RGB}{39,240,54}

\definecolor{thesisCbluedeep}{RGB}{8,104,172}
\definecolor{thesisCbluelight}{RGB}{78,179,211}
\definecolor{thesisCbluebright}{RGB}{8,104,240}

\definecolor{calibC1}{RGB}{255,255,217}
\definecolor{calibC2}{RGB}{190,238,177}
\definecolor{calibC3}{RGB}{169,233,180}
\definecolor{calibC4}{RGB}{127,205,187}
\definecolor{calibC5}{RGB}{65,182,196}
\definecolor{calibC6}{RGB}{78,179,211}
\definecolor{calibC7}{RGB}{43,140,190}
\definecolor{calibC8}{RGB}{8,104,172}
\definecolor{calibC9}{RGB}{8,64,129}
\definecolor{calibC10}{RGB}{4,24,65}

\usepackage{todonotes}
\presetkeys{todonotes}{inline}{}

\usepackage[noadjust]{cite}
\usepackage{threeparttable}
\usepackage{subcaption}
\usepackage{caption}

\IEEEoverridecommandlockouts

\newcommand{\andreas}[1]{\textcolor{red}{ANDREAS: #1}}
\newcommand{\andy}[1]{\textcolor{green}{ANDY: #1}}
\newcommand{\alex}[1]{\textcolor{teal}{ALEX: #1}}

\renewcommand{\andreas}[1]{}
\renewcommand{\andy}[1]{}
\renewcommand{\alex}[1]{}

\begin{document}

\bstctlcite{IEEEexample:BSTcontrol}

%
\title{On the Implementation Complexity of Digital Full-Duplex Self-Interference Cancellation}

%
\author{\IEEEauthorblockN{Andreas Toftegaard Kristensen\IEEEauthorrefmark{1}, Alexios Balatsoukas-Stimming\IEEEauthorrefmark{2}, and Andreas Burg\IEEEauthorrefmark{1}}
\IEEEauthorblockA{\IEEEauthorrefmark{1}Telecommunication Circuits Laboratory, \'{E}cole polytechnique f\'{e}d\'{e}rale de Lausanne, Switzerland\\
\IEEEauthorrefmark{2}Department of Electrical Engineering, Eindhoven University of Technology, Netherlands}%
\thanks{This work was supported by the Swiss NSF under grant \#200021\_182621.}%
}


\maketitle
\begin{abstract}

In-band full-duplex systems promise to further increase the throughput of wireless systems, by simultaneously transmitting and receiving on the same frequency band.
However, concurrent transmission generates a strong self-interference signal at the receiver, which requires the use of cancellation techniques.
A wide range of techniques for analog and digital self-interference cancellation have already been presented in the literature.
However, their evaluation focuses on cases where the underlying physical parameters of the full-duplex system do not vary significantly.
In this paper, we focus on adaptive digital cancellation, motivated by the fact that physical systems change over time.
We examine some of the different cancellation methods in terms of their performance and implementation complexity, considering the cost of both cancellation and training.
We then present a comparative analysis of all these methods to determine which perform better under different system performance requirements.
We demonstrate that with a neural network approach, the reduction in arithmetic complexity for the same cancellation performance relative to a state-of-the-art polynomial model is several orders of magnitude.

\end{abstract}

\andreas{%
Some points to check:
\begin{itemize}
	\item Introduction: Discuss time-varying models in the intro?
	\item Results: Add assumption on calculating the basis functions
	\item Misc: SI cancellation $\rightarrow$ SIC if we run out of space
	\item Misc: We use cancellation technique to talk about cancellation in a very general manner (also analog). We use cancellation method or canceler to refer to our methods. 
	\item Misc: We use cancellation instead of inference, perhaps define this somehwere? Although, if we use cancellation as a metric as well (or just state cancellation performance for metric?), this may cause confusion? But for the polynomial models we don't say inference, so perhaps just estimation/prediction? Make a small overview here at the top on what has to be used when
	\item Misc: Cancellation model should be reserved for our model!
	\item Misc: Consistency in saying hardware model vs dynamic hardware model?
	\item Misc: Should we say anything about the downside of the experimental and simulation approaches?
\end{itemize}}

\section{Introduction}\label{sec:introduction}

In-band full-duplex (FD) is a promising method for increasing performance by simultaneously transmitting and receiving in the same frequency band.
However, FD operation results in a strong self-interference (SI) signal at the receiver.
Fortunately, recent work has demonstrated that sufficient SI cancellation is possible to make FD systems viable~\cite{jain2011practical, duarte2012experiment, bharadia2013full, korpi2017nonlinear}.

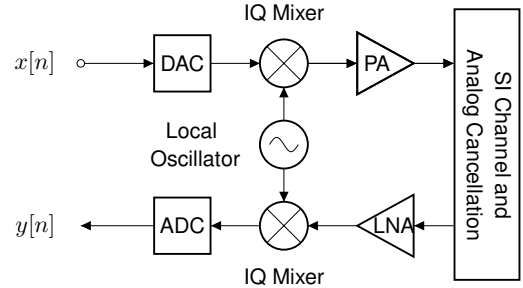
\begin{figure}[t]
	\centering
	\scalebox{0.9}{\def\cverspacing{2.4}
\def\chorzspacing{1.5}
\def\cshift{0.5}

\def\textshift{0.5}
\def\signalshift{0.25}
\def\circwidth{0.9}

\def\newscale{0.85}

\begin{circuitikz}[american, font=\sffamily, trim left=-45]

\ctikzset{blocks/scale=\newscale}

	\coordinate (coordanalogcanc) at (3*\chorzspacing,-0.5*\cverspacing);

	\coordinate (coorddac) at (0, 0);
	\coordinate (coordpa) at (2*\chorzspacing, 0);

	\coordinate (coordadc) at (0, -\cverspacing);
	\coordinate (coordlna) at (2*\chorzspacing, -\cverspacing);

	\node[mixer,scale=\newscale] (txmixer) at (\chorzspacing,0) {};

	\node[draw, rotate=-90, minimum height = 0.6*\chorzspacing cm, minimum width = 4cm, align=center, thick] (analogcanc) at (coordanalogcanc) {\small SI Channel and\\ \small Analog Cancellation};

	\draw (-\chorzspacing,0) to[short,o-] ($(coorddac) - (\cshift, 0)$)
	to[twoport,>] ($(coorddac) + (\cshift, 0)$) -- (txmixer.west) node[inputarrow]{};

	\draw (txmixer.east) to[short,-] ($(coordpa) - (\cshift, 0)$)
	to[amp,>] ($(coordpa) + (\cshift, 0)$)  -- (analogcanc.south |- 0, 0) node[inputarrow]{};

	\draw (txmixer.east) to[short,-] ($(coordpa) - (\cshift, 0)$)
	to[amp,>] ($(coordpa) + (\cshift, 0)$)  -- (analogcanc.south |- 0, 0) node[inputarrow]{};

	\node[mixer,scale=\newscale] (rxmixer) at (\chorzspacing,-\cverspacing) {};	

	\draw (analogcanc.south |- 0, -\cverspacing) -- ($(coordlna) + (\cshift, 0)$) to[amp,>] ($(coordlna) - (\cshift, 0)$) -- (rxmixer.east) node[inputarrow, rotate=180]{};

	\draw (rxmixer.west) -- ($(coordadc) + (\cshift, 0)$) to[twoport,>] ($(coordadc) - (\cshift, 0)$) -- (-\chorzspacing, -\cverspacing) node[inputarrow, rotate=180]{};

	\node[oscillator,scale=\newscale] (ref) at (0,-0.5*\cverspacing -| rxmixer.east)  {};
	\draw (ref.south) to[short,-] (rxmixer.north) node[inputarrow,rotate=270]{};
	\draw (ref.north) to[short,-] (txmixer.south) node[inputarrow,rotate=90]{};
	\draw ($(ref)+(-2*\textshift,0)$) node[left,text width=1.2cm,align=center] {\small{Local\\Oscillator}};

	\draw ($(txmixer)+(0,\textshift)$) node[above] {\small{IQ Mixer}};
	\draw ($(rxmixer)+(0,-\textshift)$) node[below] {\small{IQ Mixer}};
	\draw ($(coordpa)+(-0.125,0)$) node[] {\small{PA}};
	\draw ($(coordlna)+(0.125,0)$) node[] {\small{LNA}};

	\draw ($(coorddac)+(0,0)$) node[] {\small{DAC}};
	\draw ($(coordadc)+(0,0)$) node[] {\small{ADC}};

	\node[anchor=east] at ($(-\chorzspacing,0) + (-\signalshift, 0)$) {\normalsize $x[n]$};
	\node[anchor=east] at ($(-\chorzspacing,-\cverspacing) + (-\signalshift, 0)$) {\normalsize $y[n]$};

\end{circuitikz}

		}
	\caption{Simplified wireless FD transceiver block diagram.}
	\label{fig:system}
\end{figure}

In practice, SI cancellation is performed in multiple steps to reduce the SI signal to the level of the receiver noise floor.
As the SI signal is several orders of magnitudes larger than the signal-of-interest, the SI is first partially removed in the analog RF domain with passive and/or active suppression to avoid saturating the analog front-end of the receiver.
A residual SI signal generally remains after analog cancellation, which must be canceled in the digital domain.
This digital cancellation requires modeling the non-linear distortion and memory effects, such as digital-to-analog converter (DAC) and analog-to-digital-converter (ADC) non-linearities~\cite{balatsoukas2015baseband}, IQ imbalance~\cite{balatsoukas2015baseband, korpi2014widely}, phase-noise~\cite{sahai2013impact, syrjala2014analysis}, and power amplifier (PA) non-linearities~\cite{balatsoukas2015baseband, korpi2014widely, anttila2014modeling, korpi2017nonlinear} in addition to the memory effects of the channel as illustrated in Fig~\ref{fig:system}.
To describe and adjust the parameters of these effects various methods have been proposed such as polynomial models~\cite{kim2001digital, korpi2014widely, korpi2017nonlinear, korpi2018modeling}, support vector machines~\cite{auer2020support}, neural networks (NNs)~\cite{balatsoukas2018non, kristensen2019advanced, kristensen2020identification}, and more traditional machine learning methods~\cite{dikmese2019behavioral}.
However, while the cancellation performance and complexity of time-invariant systems have already been investigated thoroughly, we note that the parameters of FD systems change over time due to, e.g., circuit temperature variations and motion in the surrounding environment.
Consequently, models must be retrained regularly to remain accurate.
Therefore, it is critical from an implementation point-of-view to consider not only the cancellation performance and complexity, but also the tracking performance and training complexity.

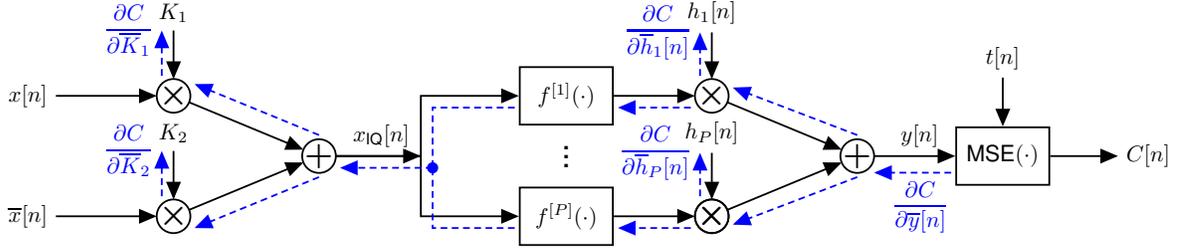
\begin{figure*}[t]
	\centering
	\scalebox{0.85}{\def\horizontalbigsep{3cm}
\def\vertbigsep{2.9cm}

\def\horizontalsmallsep{3.5cm}
\def\vertsmallsep{2cm}

\def\bpsep{0.4cm}
\def\bpsepdiag{0.283cm} 
\def\bpgap{0.1cm} 

\def\filterheight{0.9cm}
\def\filterwidth{1.456cm}

\begin{tikzpicture}[-triangle 45, scale=0.65, thick]
	\tikzstyle{dspfilter}=[rectangle, draw=black, minimum width=\filterwidth, minimum height=\filterheight, inner sep=2pt, thick]
	\tikzstyle{arithmetic}=[circle, draw=black, minimum size=15pt, inner sep=-5pt, outer sep=0pt, thick]



    \node[] (xn) at (0,0) {$x[n]$};
    \node[] (xnconj) at ($(xn) + (0,-\vertbigsep)$) {$\overline{x}[n]$};

    \node[] (k1) at ($(xn) + (\horizontalsmallsep,\vertsmallsep)$) {$K_1$};
    \node[] (k2) at ($(xnconj) + (\horizontalsmallsep,\vertsmallsep)$) {$K_2$};

    \node[arithmetic] (mult_xn_k1) at ($(xn) + (\horizontalsmallsep,0)$) {\Large $\bm{\times}$};

    \node[arithmetic] (mult_xnconj_k2) at ($(xnconj) + (\horizontalsmallsep,0)$) {\Large $\bm{\times}$};

    \node[arithmetic] (add_k1_k2) at ($(mult_xn_k1)!0.5!(mult_xnconj_k2) + (\horizontalsmallsep,0)$) {\Large $\bm{+}$};

    \coordinate[] (xiqn) at ($(add_k1_k2) + (0.7*\horizontalsmallsep,0)$) {};
    \coordinate[] (xiqn_bp) at ($(add_k1_k2) + (0.7*\horizontalsmallsep,0) + (\bpsepdiag, -\bpsepdiag)$) {};
    \node[] () at ($(add_k1_k2) + (1.5,0.5)$) {$x_{\text{IQ}}[n]$};


    \node[dspfilter] (function1) at ($(xiqn) + (\horizontalsmallsep,0.5*\vertbigsep)$) {$f^{[1]}(\cdot)$};
    \node[dspfilter] (functionP) at ($(xiqn) + (\horizontalsmallsep,-0.5*\vertbigsep)$) {$f^{[P]}(\cdot)$};

    \node[] (h1) at ($(function1) + (\horizontalsmallsep,\vertsmallsep)$) {$h_{1}[n]$};
    \node[] (hP) at ($(functionP) + (\horizontalsmallsep,\vertsmallsep)$) {$h_{P}[n]$};

    \node[] at ($(function1)!0.45!(functionP)$) {$\bm{\vdots}$};

    \node[arithmetic] (mult_f1_h1) at ($(function1) + (\horizontalsmallsep,0)$) {\Large $\bm{\times}$};

    \node[arithmetic] (mult_fP_hP) at ($(functionP) + (\horizontalsmallsep,0)$) {\Large $\bm{\times}$};

    \node[arithmetic] (mult_fP_hP) at ($(functionP) + (\horizontalsmallsep,0)$) {\Large $\bm{\times}$};

    \node[arithmetic] (add_tn) at ($(mult_f1_h1)!0.5!(mult_fP_hP) + (\horizontalsmallsep,0)$) {\Large $\bm{+}$};

    \node[] () at ($(add_tn) + (1.5,0.5)$) {$y[n]$};


    \node[dspfilter] (sub_y_t) at ($(add_tn) + (\horizontalsmallsep,0)$) {$\text{MSE}(\cdot)$};
    \node[] (target) at ($(sub_y_t.north) + (0,\vertsmallsep) + (0, -10pt)$) {$t[n]$};

    \node[] (cn) at ($(sub_y_t) + (\horizontalsmallsep,0)$) {$C[n]$};



    \draw (xn) -- (mult_xn_k1);
    \draw (k1) -- (mult_xn_k1);

    \draw (xnconj) -- (mult_xnconj_k2);
    \draw (k2) -- (mult_xnconj_k2);

    \draw (mult_xn_k1) -- (add_k1_k2);	 
    \draw (mult_xnconj_k2) -- (add_k1_k2);


    \draw (add_k1_k2) -- (xiqn);


    \draw (add_k1_k2) -- (xiqn) |- (function1);
    \draw (add_k1_k2) -- (xiqn) |- (functionP);

    \draw (h1) -- (mult_f1_h1);
    \draw (function1) -- (mult_f1_h1);

    \draw (hP) -- (mult_fP_hP);
    \draw (functionP) -- (mult_fP_hP);

    \draw (mult_f1_h1) -- (add_tn);	 
    \draw (mult_fP_hP) -- (add_tn);

    \draw (add_tn) -- (sub_y_t);


    \draw (target) -- (sub_y_t);
    \draw (sub_y_t) -- (cn);


    \draw[blue, densely dashed, densely dashed] ($(sub_y_t.west) + (-\bpgap,-\bpsep)$) -- node [below, pos=0.4] {$\dfrac{\partial C}{\partial \overline{y}[n]}$} ($(add_tn.east) + (0,-\bpsep)$);

    \draw[blue, densely dashed] ($(add_tn.north) + (0,\bpgap)$) -- ($(mult_f1_h1.east) + (\bpgap,2*\bpgap)$);
    \draw[blue, densely dashed] ($(add_tn.south) + (0,-\bpgap)$) -- ($(mult_fP_hP.east) + (\bpgap,-2*\bpgap)$);

    \draw[blue, densely dashed] ($(mult_f1_h1) + (-\bpsepdiag,\bpsepdiag) + (0, 2*\bpgap)$) -- node [left, pos=1.0] {$\dfrac{\partial C}{\partial \overline{h}_{1}[n]}$} ($(h1.south) + (-\bpsepdiag,0)$);
    \draw[blue, densely dashed] ($(mult_fP_hP) + (-\bpsepdiag,\bpsepdiag) + (0, 2*\bpgap)$) -- node [left, pos=1.0] {$\dfrac{\partial C}{\partial \overline{h}_{P}[n]}$} ($(hP.south) + (-\bpsepdiag,0)$);

    \draw[blue, densely dashed] ($(mult_f1_h1.west) + (-\bpgap,-\bpsepdiag)$) -- ($(function1.east) + (\bpgap,-\bpsepdiag)$);
    \draw[blue, densely dashed] ($(mult_fP_hP.west) + (-\bpgap,-\bpsepdiag)$) -- ($(functionP.east) + (\bpgap,-\bpsepdiag)$);

    \draw[blue, densely dashed, -] ($(function1.west) + (-\bpgap,-\bpsepdiag)$) -- ($(xiqn_bp |- function1) + (0,-\bpsepdiag)$) -- ($(xiqn_bp) + (0,0)$);
    \draw[blue, densely dashed, -] ($(functionP.west) + (-\bpgap,-\bpsepdiag)$) -- ($(xiqn_bp |- functionP) + (0,-\bpsepdiag)$) -- ($(xiqn_bp) + (0,0)$);
    \node[thick, circle, fill=blue, minimum width=0.175cm, inner sep=0] at (xiqn_bp) {};

    \draw[blue, densely dashed] (xiqn_bp) -- node [below, pos=0.6] {} ($(add_k1_k2.east) + (\bpgap,-\bpsepdiag)$);


    \draw[blue, densely dashed] ($(add_k1_k2.north) + (-0,\bpgap)$) -- ($(mult_xn_k1.east) + (\bpgap,\bpsepdiag)$);
    \draw[blue, densely dashed] ($(add_k1_k2.south) + (-0,-\bpgap)$) -- ($(mult_xnconj_k2.east) + (\bpgap,-\bpsepdiag)$);

    \draw[blue, densely dashed] ($(mult_xn_k1) + (-\bpsepdiag,\bpsepdiag) + (0, 2*\bpgap)$) -- node [left, pos=1.0] {$\dfrac{\partial C}{\partial \overline{K}_1}$} ($(k1.south) + (-\bpsepdiag,0)$);
    \draw[blue, densely dashed] ($(mult_xnconj_k2) + (-\bpsepdiag,\bpsepdiag) + (0, 2*\bpgap)$) -- node [left, pos=1.0] {$\dfrac{\partial C}{\partial \overline{K}_2}$} ($(k2.south) + (-\bpsepdiag,0)$);

\end{tikzpicture}}
	\caption{Model-based NN obtained by unfolding \eqref{eq:pa} with $f^{[P]}(z)=z|z|^{P-1}$ as the non-linearity and the MSE cost function applied at the end~\cite{kristensen2020identification}.
	Black arrows show forward propagation and blue dashed arrows show backpropagation.}
	\label{fig:mbnn}
\end{figure*}

\subsubsection*{Contribution}

In this paper, we investigate the performance and complexity of digital cancelers when tracking a time-varying hardware model of a FD transceiver.
More specifically, we apply polynomial and model-based NN cancelers at various adaptation rates to investigate their cancellation/complexity trade-offs.
The exploration shows that using backpropagation for tracking, as done for our model-based NN, can provide orders of magnitudes reduction in arithmetic complexity for the same cancellation performance relative to the other methods.
To the best of our knowledge, this is the first work that considers the complexity of FD cancellation methods when adaptively tracking a time-varying system.

\section{Dynamic Hardware Model}\label{sec:dynamic_hardware_model}

In this section, we first describe how we model hardware non-linearities and the propagation channel of a FD transceiver for the purpose of this study.
Then, we describe how to model the time-varying dynamics of this hardware model.

\subsection{Non-Linearities of Full-Duplex Transceivers}

A FD transceiver contains many components.
However, to evaluate SI cancellation performance, we only consider the well-understood non-linearities and the propagation channel for our model.
Fig.~\ref{fig:system} shows a simplified block diagram of the FD transceiver under consideration, with the main non-linearities being the IQ-mixer and the PA in the transmitter~\cite{korpi2017nonlinear}.
Furthermore, we assume that there is no signal-of-interest present from a remote node and, therefore, the received signal $y[n]$ is the SI signal that needs to be modeled and then canceled in the digital domain.

The output $\hat{y}[n]$ of the PA can be modeled using a memory polynomial, with the output of the IQ mixer as input~\cite{korpi2017nonlinear}
\begin{equation}\label{eq:pa}
	\hat{y}[n] = \sum_{m=0}^{M-1} \sum _{\substack{p=1\\p \text{ odd}}}^P h_{p} [m] x_{\text{IQ}}[n-m]|x_{\text{IQ}}[n-m]|^{p-1} \, ,
\end{equation}
where $x_{\text{IQ}}[n] = K_1 x[n] + K_2\overline{x}[n]$ represents the effect of the IQ-mixer on the input signal $x[n]$, $M$ is the overall memory of the system, and $P$ is the non-linearity order.
This model also accounts for the propagation channel and any analog SI cancellation, which merely attenuates and adds delayed copies of the transmitted signal to the overall received signal.

\subsection{Time-Varying Dynamics of Full-Duplex Transceivers}

With~\eqref{eq:pa} as a hardware model that accounts for the propagation channel and the non-linearities, we now consider how to model the time-varying dynamics of this model.
To this end, we consider each hardware model parameter to be sampled from a stochastic process, described by an autoregressive (AR) model.
The notation AR(1) indicates an autoregressive model of order one.
A complex-valued AR(1)-process can be written as follows for a complex random variable $W$ at time-step $t$
\begin{equation}\label{eq:ar}
	W_t = c + \beta W_{t-1} + \epsilon_t \, ,
\end{equation}
where $\epsilon \sim \mathcal{CN}(0, \sigma^2_{\epsilon})$, $c$ a complex-valued parameter, and $\beta$ a real-valued parameter of the process.\footnote{While $\beta$ is generally complex for a complex-valued AR(1)-process, we assume it is real for the sake of simplicity.}

The parameter $\beta$ is the main parameter of interest in this paper. 
As an AR(1)-process is an IIR filter with noise as input, the $\beta$ defines the cut-off frequency of this IIR filter.
A value of $\beta$ close to 1 results in significant attenuation of the noise contribution and the long-term stochastic trend dominates.
On the other hand, a small value of $\beta$ results in little attenuation of the noise and faster short-term trends dominate.
With this in mind, we can take two views as to what $\beta$ means for our setup.
The first view considers $\beta$ as defining the physical rate of change of a process in terms of its cut-off frequency and changing $\beta$ will change the physical rate of change. 
The second view considers $\beta$ as the adaptation rate to a process with a fixed underlying physical rate of change and changing $\beta$ means that the adaptation rate to the process is changed.
As this paper focuses on adaptive digital SI cancellation, we assume the latter.
Hence, we start from a fixed physical rate of change and adjust the adaptation rate $\beta$ to analyze the tracking performance and complexity for different adaptation rates.

The process is wide-sense stationary if $\beta < 1$.
Given wide-sense stationarity, the mean and variance of the random variables are given as  $\mathrm{E}\{W\} = \frac{c}{1-\beta}$ and $\mathrm{Var}\{W\} = \frac{\sigma^2_{\epsilon}}{1-\beta^2}$, respectively.
Then, if we define $\beta$ and the expectations and variances of our parameters, AR(1)-processes can be defined for each parameter and we can sample from these processes to generate a population of time-varying hardware models.

\section{Cancellation Methods}\label{sec:cancellation_methods}

In this section, we describe the three cancellation methods that we consider.
We first specify the polynomial models and then we describe the NN model.

\subsection{Polynomial Models}

As a base-line model for our evaluation, we use  a linear model given as
\begin{equation}\label{eq:lin}
	\hat{y}[n] = \sum_{m=0}^{M-1} h[m] x[n-m] \, ,
\end{equation}
which is equivalent to~\eqref{eq:pa} with $P=1$ and $K_2 = 0$.
While this model only accounts for the SI channel and memory effects from the transceiver chain and ignores all non-linearities, it only has $M$ parameters and is therefore expected to track reasonably well for a large range of adaptation rates.

To model the non-linearities, we consider the widely-linear memory polynomial (WLMP), which is similar to~\eqref{eq:pa}, but it is linear in its parameters so it can be fitted directly with least squares (LS) methods~\cite{korpi2017nonlinear}.
The WLMP is given as
\begin{equation}\label{eq:wlmp}
	y_{\text{SI}}[n] = \sum _{\substack{p=1,\\p \text{ odd}}}^P \sum_{q=0}^p\sum_{m=0}^{M-1} g_{p,q}[m] x[n-m]^{q}\overline{x}[n-m]^{p-q} \, ,
\end{equation}
where the parameters $g_{p,q}$ contain the combined effects from the IQ-mixer, PA, and the SI channel.
This polynomial model has $\frac{M}{4}(P+1)(P+3)$ parameters, so its complexity scales with $MP^2$ which may make tracking harder.

\subsection{Model-Based Neural Network}

Instead of generalizing~\eqref{eq:pa} to obtain an over-parameterized generic polynomial model, as in~\eqref{eq:wlmp}, we can also consider directly expanding~\eqref{eq:pa} with $x_{\text{IQ}}$ to keep the number of parameters at $N_p = \frac{1}{2}(P+1)M+2$.
This can be achieved by unfolding~\eqref{eq:pa} to obtain a computational graph representation of~\eqref{eq:pa}, as shown in Fig.~\ref{fig:mbnn}.
We refer to this representation as a model-based NN (MBNN).
However, this model is not linear in its parameters, so it cannot be directly fitted using LS methods.
Instead, backpropagation is used to obtain the gradients of each parameter in a computationally efficient manner, and gradient descent methods are used to optimize these parameters~\cite{kristensen2020identification}.
This model directly matches the blocks of our hardware model and thus learns the physical parameters in a more direct manner.
Additionally, backpropagation is a computationally efficient method to calculate the gradients by avoiding duplicate and other unnecessary calculations.

\section{Experimental Setup}

%


%

In this section, we describe the dataset on which we train our polynomial and NN models.
First, we describe the assumptions we make on the distributions of the parameters as the dataset is generated using the hardware model.
Then, we describe how the dataset is created using the hardware model and how we evaluate the tracking performance of the three cancellation methods under consideration on this dataset.

\subsection{Hardware Model Parameters}

The digital cancelers are trained on a synthetic dataset of baseband samples created by feeding transmit baseband samples to the dynamic hardware model in~\eqref{eq:pa} and sampling its outputs.
As the hardware model uses AR(1)-processes to describe the time-varying dynamics of its parameters, we have to define the distribution of each hardware model parameter and the values of $\beta$ that we consider to calculate the noise variance and $c$ for the AR(1)-process in~\eqref{eq:ar}.

We define the parameters $K_1$ and $K_2$ of the IQ-mixer as $K_1 = \frac{1}{2}(1+A_{\text{IQ}}e^{-j \phi_{\text{IQ}}})$ and $K_2 = \frac{1}{2}(1-A_{\text{IQ}}e^{j \phi_{\text{IQ}}})$, where $A_{\text{IQ}}$ and $\phi_{\text{IQ}}$ are the gain and phase imbalance parameters, respectively.
The parameters $A_{\text{IQ}}$ and $\phi_{\text{IQ}}$ are then the parameters for which we define AR(1)-processes.
We use the image-rejection ratio (IRR), given as $10\log_{10}\frac{|K_1|^2}{|K_2|^2}$, to help define their distributions.
Generally, the IRR lies in the range of \SI{20}{\decibel} to \SI{40}{\decibel}~\cite{valkama2002signal, anttila2008circularity, anttila2014modeling, komatsu2020iterative}.
The mean of $A_{\text{IQ}}$ and $\phi_{\text{IQ}}$ are set to \num{1} and \num{0}, respectively, and the variance of $A_{\text{IQ}}$ and $\phi_{\text{IQ}}$ are set to \num{0.005}.
These values result in \SI{95}{\percent} of the samples having an IRR lying in the range of \SI{20}{\decibel} to \SI{40}{\decibel}.

For the parameters of the memory polynomial model, we also have to define a distribution for each parameter and we have to define $M$ and $P$.
Since an SI signal is observed as a strong LOS component at the receiver, taps at $m=0$ are defined as Rice distributed, and the remaining taps as Rayleigh distributed.
In~\cite{kurzo2020hardware} we experimentally determined that a relatively short channel is sufficient, and, therefore, we set $M=3$ and $P=5$ for the memory polynomial.
Then, by fitting~\eqref{eq:pa} on the dataset in~\cite{kurzo2020hardware} using LS, assuming $K_1=1$ and $K_2=0$, we obtain estimates of the parameters $h_p[m]$ in our testbed.
We use the power of these to help define the distribution parameters.
The average reduction in power from a parameter $h_p[m]$ to $h_{p+1}[m]$ and $h_p[m+1]$ is found through measurements to be around \SI{20}{\decibel}.
We use this \SI{20}{\decibel} reduction to define the power of the taps, that is, if the power of $h_1[0]$ is \SI{0}{\decibel}, the power of $h_3[0]$ and $h_1[1]$ is \SI{-20}{\decibel} and the power of $h_5[0]$ is \SI{-40}{\decibel}.

Finally, we require $\beta$, which defines the rate at which our cancellation methods are adapting to changes in the hardware.
We investigate $\beta \in \{0.9, 0.99, \dots, 0.99999\}$, corresponding to oversampling rates from \num{1}$\times$ to \num{10000}$\times$.


\subsection{Dataset}

We evaluate our cancellation methods on 250 datasets, using a set of 50 different seed values, for each of the 5 values of $\beta$.
The assumptions on the hardware model parameters from the previous section are then used to define AR(1)-processes for each parameter for each value of $\beta$ by adjusting $c$ and the noise variance accordingly.

Each dataset contains a period in which the hardware is stable and one in which it changes.
We refer to the first period as the static period and the period where the hardware changes as the dynamic period.
There are \num{10000} samples generated from the static period for the canceler to converge and then \num{10000} samples for evaluating the tracking on the dynamic period.
As input to the hardware model when generating these datasets, we use an OFDM frame of \num{20000} baseband samples split into half for the static and dynamic periods.

For the static period, we then generate 50 AR(1)-processes for each hardware model parameter which gives 50 model realizations using the first sample of each AR(1)-process.
For the dynamic period, we evaluate all AR(1)-processes for \num{10000} steps for each $\beta$ using the first sample from the static period as the initial value.

To fit our polynomial models on these datasets, we use LS for the initial fit on the static period.
For the dynamic period, we use least mean squares (LMS) for both the linear canceller and the WLMP.
Additionally, recursive least squares (RLS) is used for the WLMP on the dynamic period.
For the  MBNN, the FTRL optimizer is used for both the static and dynamic periods.
The MBNN is trained for 5 epochs on the static period to get a good initial performance.

To evaluate the performance of our cancellation methods on the datasets, we use cancellation defined as $C_{\text{dB}} = 10 \log_{10} \left( \frac{\sum_n |t[n]|^2}{\sum_n |t[n]-y[n]|^2 } \right)$, where $t[n]$ is the received SI signal target at time-step $n$ and $y[n]$ our corresponding estimation.
To help prevent over-fitting, we add Gaussian noise at \SI{-40}{\decibel} to the generated output values.
Furthermore, to determine the best learning parameters for each model for each value of $\beta$, we reserve the first 10 seeds to find the learning parameter which gives the highest average cancellation for each $\beta$ on the dynamic period.
The remaining datasets are used for the results.

\section{Results}\label{sec:results}

In this section, we first show the cancellation performance as a function of $\beta$ and then we consider the computational complexity as a function of the cancellation performance.

\begin{figure}[t]
	\centering
	{\begin{tikzpicture}
    \begin{axis}[
		normalsize,
		width = 5.6385cm,
		height = 3.0cm,
		scale only axis,
		xmin=0, xmax=4,
		ymin=12, ymax=40,
		ymajorgrids=true,
		grid style=dashed,
		xlabel = {Oversampling Rate},
		ylabel = {Cancellation (dB)},
		ylabel near ticks,
		xlabel near ticks,
	    xtick={0, 1, 2, 3, 4},
	    ytick={15, 20, 25, 30, 35, 40},
		xticklabels = {{\SI{1}{}$\times$}, {\SI{10}{}$\times$}, {\SI{100}{}$\times$}, {\SI{1000}{}$\times$}, {\SI{10000}{}$\times$}},
	    ytick distance = 5,
		label style={font=\small},
		tick label style={font=\small},
		ymajorgrids,
		legend style={at={(0.5,1.4)}, anchor=north, font=\footnotesize},
		legend columns=2,
		legend cell align=left,
    ]

	\def\dataset{figures/results_summary/model_based_nn_ftrl_canc_dynamic_total_all_ls_std.csv}

	\addplot[thesisCbluedeep, very thick, solid, mark=*] table[x index = 0, y index = 1] {\dataset};
    \addlegendentry{MBNN - FTRL}

	\addplot+[name path=test_mbnnbot, color=thesisCbluedeep, mark=none, forget plot] table[x index = 0, y index = 2] {\dataset};
	\addplot+[name path=test_mbnntop, color=thesisCbluedeep, mark=none, forget plot] table[x index = 0, y index = 3] {\dataset};
	\addplot[thesisCbluedeep!50,fill opacity=0.5, forget plot] fill between[of=test_mbnnbot and test_mbnntop];

	\def\dataset{figures/results_summary/wlmp_rls_canc_dynamic_total_all_ls_std.csv}

	\addplot[thesisCreddeep, very thick, solid, mark=square*] table[x index = 0, y index = 1] {\dataset};
    \addlegendentry{WLMP - RLS}

	\addplot+[name path=test_wlmpbot, color=thesisCreddeep, mark=none, forget plot] table[x index = 0, y index = 2] {\dataset};
	\addplot+[name path=test_wlmptop, color=thesisCreddeep, mark=none, forget plot] table[x index = 0, y index = 3] {\dataset};
	\addplot[thesisCreddeep!50,fill opacity=0.5, forget plot] fill between[of=test_wlmpbot and test_wlmptop];

	\def\dataset{figures/results_summary/wlmp_lms_canc_dynamic_total_all_ls_std.csv}

	\addplot[thesisCorangedeep, very thick, solid, mark=diamond*] table[x index = 0, y index = 1] {\dataset};
    \addlegendentry{WLMP - LMS}

	\addplot+[name path=test_wlmpbot, color=thesisCorangedeep, mark=none, forget plot] table[x index = 0, y index = 2] {\dataset};
	\addplot+[name path=test_wlmptop, color=thesisCorangedeep, mark=none, forget plot] table[x index = 0, y index = 3] {\dataset};
	\addplot[thesisCorangedeep!50,fill opacity=0.5, forget plot] fill between[of=test_wlmpbot and test_wlmptop];

	\def\dataset{figures/results_summary/lin_lms_canc_dynamic_total_all_ls_std.csv}

	\addplot[thesisCgreendeep, very thick, solid, mark=triangle*] table[x index = 0, y index = 1] {\dataset};
    \addlegendentry{Linear - LMS}

	\addplot+[name path=test_linbot, color=thesisCgreendeep, mark=none, forget plot] table[x index = 0, y index = 2] {\dataset};
	\addplot+[name path=test_lintop, color=thesisCgreendeep, mark=none, forget plot] table[x index = 0, y index = 3] {\dataset};
	\addplot[thesisCgreendeep!50,fill opacity=0.5, forget plot] fill between[of=test_linbot and test_lintop];

    \end{axis}
\end{tikzpicture}}
	\caption{Average cancellation on the dynamic period $\pm 1$ standard deviation as a function of oversampling relative to the physical rate of change.}
	\label{fig:canc_dynamic_total_all_ls_std}
\end{figure}
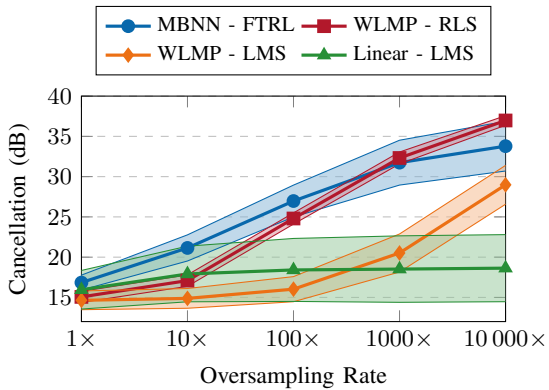

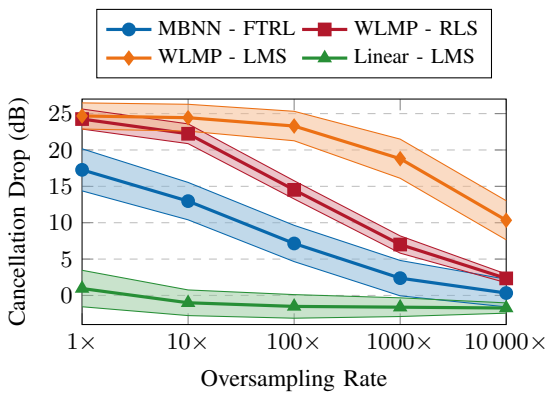
\begin{figure}[t]
	\centering
	{\begin{tikzpicture}
    \begin{axis}[
		normalsize,
		width = 5.6385cm,
		height = 3.0cm,
		scale only axis,
		xmin=0, xmax=4,
		ymin=-4, ymax=27,
		ymajorgrids=true,
		grid style=dashed,
		xlabel = {Oversampling Rate},
		ylabel = {Cancellation Drop (dB)},
		ylabel near ticks,
		xlabel near ticks,
	    xtick={0, 1, 2, 3, 4},
		xticklabels = {{\SI{1}{}$\times$}, {\SI{10}{}$\times$}, {\SI{100}{}$\times$}, {\SI{1000}{}$\times$}, {\SI{10000}{}$\times$}},
	    ytick distance = 5,
		label style={font=\small},
		tick label style={font=\small},
		ymajorgrids,
		legend style={at={(0.5,1.4)}, anchor=north, font=\footnotesize},
		legend columns=2,
		legend cell align=left,
    ]

	\def\dataset{figures/results_summary/model_based_nn_ftrl_canc_drop_all_ls_std.csv}

	\addplot[thesisCbluedeep, very thick, solid, mark=*] table[x index = 0, y index = 1] {\dataset};
    \addlegendentry{MBNN - FTRL}

	\addplot+[name path=test_mbnnbot, color=thesisCbluedeep, mark=none, forget plot] table[x index = 0, y index = 2] {\dataset};
	\addplot+[name path=test_mbnntop, color=thesisCbluedeep, mark=none, forget plot] table[x index = 0, y index = 3] {\dataset};
	\addplot[thesisCbluedeep!50,fill opacity=0.5, forget plot] fill between[of=test_mbnnbot and test_mbnntop];

	\def\dataset{figures/results_summary/wlmp_rls_canc_drop_all_ls_std.csv}

	\addplot[thesisCreddeep, very thick, solid, mark=square*] table[x index = 0, y index = 1] {\dataset};
    \addlegendentry{WLMP - RLS}

	\addplot+[name path=test_wlmpbot, color=thesisCreddeep, mark=none, forget plot] table[x index = 0, y index = 2] {\dataset};
	\addplot+[name path=test_wlmptop, color=thesisCreddeep, mark=none, forget plot] table[x index = 0, y index = 3] {\dataset};
	\addplot[thesisCreddeep!50,fill opacity=0.5, forget plot] fill between[of=test_wlmpbot and test_wlmptop];

	\def\dataset{figures/results_summary/wlmp_lms_canc_drop_all_ls_std.csv}

	\addplot[thesisCorangedeep, very thick, solid, mark=diamond*] table[x index = 0, y index = 1] {\dataset};
    \addlegendentry{WLMP - LMS}

	\addplot+[name path=test_wlmpbot, color=thesisCorangedeep, mark=none, forget plot] table[x index = 0, y index = 2] {\dataset};
	\addplot+[name path=test_wlmptop, color=thesisCorangedeep, mark=none, forget plot] table[x index = 0, y index = 3] {\dataset};
	\addplot[thesisCorangedeep!50,fill opacity=0.5, forget plot] fill between[of=test_wlmpbot and test_wlmptop];

	\def\dataset{figures/results_summary/lin_lms_canc_drop_all_ls_std.csv}

	\addplot[thesisCgreendeep, very thick, solid, mark=triangle*] table[x index = 0, y index = 1] {\dataset};
    \addlegendentry{Linear - LMS}

	\addplot+[name path=test_linbot, color=thesisCgreendeep, mark=none, forget plot] table[x index = 0, y index = 2] {\dataset};
	\addplot+[name path=test_lintop, color=thesisCgreendeep, mark=none, forget plot] table[x index = 0, y index = 3] {\dataset};
	\addplot[thesisCgreendeep!50,fill opacity=0.5, forget plot] fill between[of=test_linbot and test_lintop];

    \end{axis}
\end{tikzpicture}}
	\caption{Average cancellation drop between the static and dynamic periods $\pm 1$ standard deviation as a function of oversampling relative to the physical rate of change.}
	\label{fig:canc_drop_total_all_ls_std}
\end{figure}

\subsection{Dynamic Cancellation}

In Fig.~\ref{fig:canc_dynamic_total_all_ls_std}, we show the cancellation performance $\pm1$ standard deviation over the different datasets on the dynamic period as a function of the oversampling rate relative to the physical rate of change.

We observe that at an oversampling rate of \num{1}$\times$, all models converge towards the same poor performance and slowly improve their cancellation quality with a larger oversampling rate.
As an oversampling rate of \num{10000}$\times$ is necessary for the models to start tracking with their initial performance from the static period, we conclude that, in general, significant oversampling is required to maintain peak cancellation quality.

In terms of peak cancellation quality, the WLMP using RLS achieves the best cancellation, although the MBNN is similar or better for oversampling rates below \num{1000}$\times$  compared to the WLMP and linear cancelers.
Moreover, the WLMP using RLS shows less performance variation across oversampling rates and is thus a better guarantee for good performance at high oversampling rates if its high computational cost is acceptable.

\subsection{Cancellation Drop}

In Fig.~\ref{fig:canc_dynamic_total_all_ls_std}, we calculate the reduction in average cancellation quality between the static and dynamic periods as a function of the oversampling rate.
We observe that the linear canceler improves over its static performance slightly at \num{10}$\times$ oversampling rate, indicating that the linear canceler can adapt so quickly that it can even follow changes in the hardware that allow for a quality improvement.
For the WLMP with RLS, even at \num{10000}$\times$ the cancellation quality is \SI{2.5}{\decibel} below the static fit, confirming the previous observation that significant oversampling is required.
We also observe that LMS is not suitable for WLMP as we observe a \SI{10}{\decibel} reduction at even \SI{10000}{}$\times$ oversampling.
Furthermore, the MBNN proves to be more robust to lower oversampling compared to WLMP using RLS, considering the large gap between the MBNN and the WLMP using RLS for the first few oversampling rates.

\begin{table}[t]
	\centering
	\small
	\caption{Number of real-valued parameters and the arithmetic complexity for training each model.}
	\label{tab:complexity}
	\begin{tabular}{l|rrrr}
		~	& \multicolumn{1}{c}{Lin} 	& \multicolumn{1}{c}{WLMP - LMS} & \multicolumn{1}{c}{WLMP - RLS} & \multicolumn{1}{c}{MBNN} \\
		\hline
		N\textsubscript{params}	    & \num{6}	& \num{72} & \num{72} & \num{22} \\
		N\textsubscript{add}		& \num{47}	& \num{509} & \num{34668} & \num{657} \\
		N\textsubscript{mult}		& \num{21}	& \num{219} & \num{16092} & \num{391} \\
		N\textsubscript{div}		& \num{0}	& \num{0} & \num{72} & \num{40} \\
		N\textsubscript{sqrt}		& \num{0}	& \num{0} & \num{0} & \num{22} \\
	\end{tabular}
\end{table}


\subsection{Model Complexity}

Finally, we consider the computational and memory complexity of the models.
In Table~\ref{tab:complexity}, we provide the real-valued complexity in terms of the number of parameters, additions, multiplications, divisions, and square-root operations for the various SI cancelers.
The arithmetic complexity represents the combined cost of a single prediction and parameter update, including the cost of calculating all the gradients of the MBNN.
Complex-valued additions and multiplications are converted to real-valued operations assuming that one complex multiplication can be implemented using three real multiplications and five real additions and one complex addition can be implemented using two real additions.
Complex-valued divisions are converted to real-valued divisions and multiplications by multiplying with the conjugate of the denominator, leaving a real-valued denominator.
The square-root operations for the MBNN are only for the individual real-valued parameters of the MBNN and come from the FTRL optimizer.

\begin{figure}[t]
	\centering
	{\begin{tikzpicture}
    \begin{semilogyaxis}[
		normalsize,
		width = 5.6385cm,
		height = 3.0cm,
        scale only axis,
		ymajorgrids=true,
		grid style=dashed,
		xlabel = {Cancellation (dB)},
		ylabel = {FLOPS},
		ylabel near ticks,
		xlabel near ticks,
	    xtick={15, 20, 25, 30, 35, 40},
	    ytick={100, 10000, 1000000, 100000000},
		label style={font=\small},
		tick label style={font=\small},
		ymajorgrids,
		legend style={at={(0.5,1.4)}, anchor=north, font=\footnotesize},
		legend columns=2,
		legend cell align=left,
    ]

	\def\dataset{figures/results_complexity/model_based_nn_ftrl_canc_dynamic_total_all_ls_std_flops.csv}

	\addplot[thesisCbluedeep, very thick, solid, mark=*] table[x index = 0, y index = 1] {\dataset};
    \addlegendentry{MBNN - FTRL}

	\def\dataset{figures/results_complexity/wlmp_rls_canc_dynamic_total_all_ls_std_flops.csv}

	\addplot[thesisCreddeep, very thick, solid, mark=square*] table[x index = 0, y index = 1] {\dataset};
    \addlegendentry{WLMP - RLS}

	\def\dataset{figures/results_complexity/wlmp_lms_canc_dynamic_total_all_ls_std_flops.csv}

	\addplot[thesisCorangedeep, very thick, solid, mark=diamond*] table[x index = 0, y index = 1] {\dataset};
    \addlegendentry{WLMP - LMS}

	\def\dataset{figures/results_complexity/lin_lms_canc_dynamic_total_all_ls_std_flops.csv}

	\addplot[thesisCgreendeep, very thick, solid, mark=triangle*] table[x index = 0, y index = 1] {\dataset};
    \addlegendentry{Linear - LMS}

    \end{semilogyaxis}
\end{tikzpicture}}
	\caption{Number of floating-point operations per second (FLOPS) as a function of the average cancellation performance on the dynamic period.}
	\label{fig:canc_dynamic_total_all_ls_flops}
\end{figure}
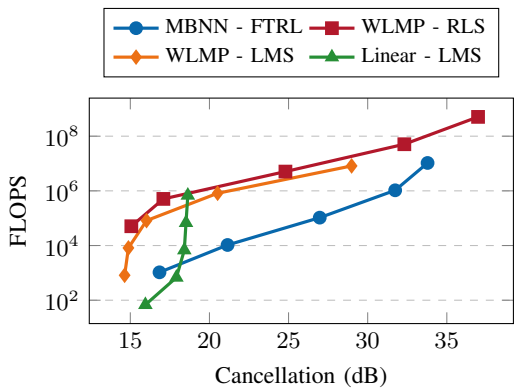

Using the total number of additions and multiplications for each model in Table~\ref{tab:complexity} as the cost in floating-point operations, we can multiply with the oversampling rates to generate Fig.~\ref{fig:canc_dynamic_total_all_ls_flops} to show the number of floating-point operations per second (FLOPS) as a function of the average cancellation on the dynamic period.
Here, we see that, except for the lowest performance requirements, where the linear canceler is best, and the highest performance requirement that only the WLMP with RLS can achieve, the MBNN is the best option in terms of FLOPS, reducing the number of FLOPS by around $10^3\times$ to $10\times$ relative to the WLMP using RLS.
While this evaluation does not consider the cost of divisions and square-root operations, the MBNN and WLMP are separated by orders of magnitudes and their cost in division and square-root operations is relatively close. 
Therefore, the MBNN shows good performance, not only in terms of cancellation quality in a static scenario, but also for tracking.
However, we still have to note that the MBNN has a higher performance variation relative to the WLMP using RLS.
If this is acceptable the MBNN represents a good trade-off relative to the WLMP.

\section{Conclusion}\label{sec:conclusion}

In this paper, we provided an investigation into the performance and complexity of digital self-interference cancelers when tracking a time-varying hardware model of a full-duplex transceiver at various oversampling rates.
We showed that using a model-based neural network for tracking provides cancellation which is either better or close to that of a state-of-the-art polynomial model.
Moreover, the model-based neural network provides a significant reduction in arithmetic complexity for a wide range of tracking quality requirements compared to other models, showing that backpropagation-based methods can be very efficient for adaptive full-duplex self-interference cancellation.
Neural network methods thus provide a good cancellation/complexity trade-off relative to more simple/complex methods.


\bibliographystyle{IEEEtran}
\bibliography{IEEEabrv,bibliography}


\end{document}